\documentclass[10pt]{revtex4}
\pdfoutput=1

\usepackage[english]{babel}
\usepackage{graphicx}
\usepackage{amssymb}
\usepackage{amsmath}
\usepackage{soul}
\usepackage{color}
\def\laq{~\raise 0.4ex\hbox{$<$}\kern -0.8em\lower 0.62ex\hbox{$\sim$}~}
\def\gaq{~\raise 0.4ex\hbox{$>$}\kern -0.7em\lower 0.62ex\hbox{$\sim$}~}

\def\beq{\begin{equation}}
\def\eeq{\end{equation}}
\def\bea{\begin{eqnarray}}
\def\eea{\end{eqnarray}}
\def\bean{\begin{eqnarray*}}
\def\eean{\end{eqnarray*}}

\begin{document}

\begin{center}
{\huge \bf  Cosmic acceleration and $f( R)$ theory: perturbed solution in a matter FLRW model}\\

\vspace{10mm}

\vspace{3mm}
{L. Cosmai$^1$, G. Fanizza$^{1,2,3}$ and L. Tedesco$^{1,2}$} \\
\vspace{6mm}
{\sl ${ }^1$Istituto Nazionale di Fisica Nucleare, Sezione di Bari, Bari, Italy}\\
\vspace{3mm}
{\sl ${ }^2$Dipartimento di Fisica, Universit\`a di Bari, Via G. Amendola 173\\70126 Bari, Italy}\\
\vspace{3mm}
{\sl${ }^3$Universite de Geneve, Departement de Physique Theorique,\\ 24 quai Ernest-Ansermet, CH-1211 Geneve 4, Switzerland}
\end{center}
\begin{abstract}
In the present paper we consider $f(R)$ gravity theories in the metric approach and we derive the equations of motion, focusing also on the boundary conditions. In such a way we apply the general equations to a first order perturbation expansion of the Lagrangian. We present  a model able to fit supernovae data without introducing dark energy.
\end{abstract}
%
%PACS numbers: 04.25Nx, 04.50.+h
%
\maketitle
\section{Introduction}
Gravity is supposed to be the only dominant force at large distances during the present epoch. According to this and due to several puzzles concerning the evolution of the Universe (for instance dark energy, dark matter and so on), it is reasonable to consider that we might have not fully understood it on a cosmological scale.
\\
The first attempts to modify the Einstein's gravity go back in 1919, when Weyl \cite{weyl} added a quadratic term in the Weyl tensor to the Einstein-Hilbert Lagrangian. Later many authors gave attention to modifications of the gravitational theory, for example Eddington \cite{eddington}, Lanczos, Bach, Schrodinger and then  Buchdahl \cite{buchdal} that analyzed the actions considering singularity free oscillating cosmology.
In the 1960's in the context of quantum gravity the Einstein's Lagrangian was modified introducing terms containing higher orders of the scalar curvature \cite{dewitt}.
\\
Very interesting classes of extended gravity are the so called ``$f(R)$ theories'',  coming from a direct generalization of the Einstein's Lagrangian (for complete reviews see 
\cite{{Amendola:2006we},{silvestri},{faraoni1},{capozziello1},{defelice1},{capozziello2},{sotiriou},{defelice},{nojiri0}}  and references therein). 
Among these, $f(R)$ gravity seems to be an interesting  model that is relatively  simple and it may have many applications in astrophysics, cosmology and high energy phenomena \cite{{elizalde},{Capozziello:2007ec}}. The paradigm consists of adding higher order curvature invariants.
The simplest modification of gravity which still preserves all the symmetries of  General Relativity (GR) consists of the extension of Einstein-Hilbert action
\begin{equation}
S_{{EH}}=-\frac{1}{16\pi\,G}\int_{\Omega} d^4x\,\sqrt{-g}\,R
\end{equation}
where instead of $R$, the Ricci scalar curvature, an arbitrary function $f(R)$ is present,
\begin{equation}
S=\int_{\Omega} d^4x\,\sqrt{-g}\,\;f(R) \, 
\end{equation}
where $g$ is determinant of the metric $g_{\mu \nu}$. From a conceptual point of view we have no a priori reason to consider the gravitational Lagrangian as a linear function of Ricci scalar R.
From the technical point of view, this way to proceed directly allows us to write field equations in order to compare them with GR ones. Moreover, they are directly related to scalar-tensor theories by a peculiar conformal transformation of the metric involving a scalar field $\phi$ \cite{chiba}. This way to proceed gives field equations which are also ghost free \cite{{Biswas:2011ar},{Biswas:2013cha}}.
\\
It has been established that our Universe is undergoing  an accelerated phase. Indeed a series of observations based on Supernovae type Ia  \cite{{riess},{perlmutter},{tonry}} can be explained by the accelerated expansion of the Universe. Within the mathematical framework of the GR and the idea that our cosmo is homogeneous and isotropic, the scientists assume that this acceleration is due to some kind of negative pressure form due to ``dark energy''.  Cosmologists have proposed many models of dark energy but those models have many free parameters and constraints from observational data. 
This discovery has revolutionized modern cosmology. There are many explanations and theoretical models in literature, for example quintessence, k-essence, Chaplygin gas... and so on. 
The simple explanation for the Universe's accelerated expansion is a cosmological constant, $\Lambda$, that is to say a nonzero vacuum energy that drives the acceleration of the Universe. From a observational viewpoint it is important to say that $\Lambda$CDM model of the Universe is in agreement with data coming from observations. But this model shows incongruences and is ``unnatural'', because poses  important theoretical questions: what is the reason why the nonzero vacuum energy should drive the acceleration?  Why is the cosmological constant so small? This  is known as  ``cosmological constant problem''  and it is a fundamental problem in cosmology and physics.  In other terms the nature of dark energy as well as its cosmological origin remain unknown and a real mystery. 
\\
For a resolution to this problem we invoke the class of $f(R)$ theories. For sure, this is just a particular class of extended gravity theories. Some different interesting approaches have been studied in \cite{Biswas:2011ar,Biswas:2013cha}.
Nevertheless recent research has shown a plausible alternative to this picture, in fact it has been shown that such cases lead to an effective dark energy 
\cite{{capozziello2},{uno},{due},{tre},{quattro},{cinque},{sei},{sette},{otto},{nove},{dieci},{undici},{dodici},{tredici},{quattordici},{quindici},{sedici},{diciassette},{diciotto},{diciannove},{venti},{ventuno},{ventidue},{ventitre},{ventiquattro},{venticinque},{ventisei},{ventisette},{ventotto},{ventinove},{trenta},{trentuno},{trentadue},{trentatre},{trentaquattro},{trentacinque},{trentasei},{trentasette},{trentotto},{trentanove},{quaranta},{cento}}; in other terms these models mimic the accelerated expansion of the Universe by a modification of general relativity that converts the attractive gravity into a repulsive interaction on cosmological scales.  
If we consider a small correction to the Einstein-Hilbert action, for example, by adding an $1/R$ term, we have an acceleration of the Universe  because of the $1/R$ term which is able to dominate as the Hubble parameter decreases. This theory shows \cite{chiba} that there is an equivalence with scalar-tensor gravity without scalar kinetic term. It is important to say that this connection to scalar-tensor gravity is provided by a conformal transformation connecting the Einstein frame and the Jordan one and is valid for all extended theories of gravity that have an action with $f(R)$, a function of Ricci scalar in which $f(R)$ has nonzero second derivate with respect to $R$. These $f(R)$ models contain higher order gravity terms that may be the cause of the acceleration of the Universe.
\\
In other terms, modifying general relativity allows us to eliminate the dark energy, but this approach does not explain the minuscule value of vacuum energy.  It is also important to stress that we don't know the exact functional form of the Lagrangian, therefore it is necessary to test theoretical considerations with observational data. 
\\
This paper is organized as follows: in Sec. II we report the well-know derivation of the modified field equations in $f(R)$ gravity; in Sec. III we focus on perturbed Lagrangian; in Sec. IV we consider the Hubble parameter; in Sec. V we study our model in relation to the apparent acceleration of the Universe. We summarize our conclusions in Sec. VI.
%
%
%%%%%%%%%%%%      SECTION 2  %%%%%%%%%%%
%
\vskip 1.3truecm
\section{Deriving field equations in $f( R)$ gravity}
In this Section we report the standard way \cite{Dyer:2008hb,Guarnizo:2010xr} to obtain the modified equations in $f(R)$ gravity. We start from a theory described by the Lagrangian density $\sqrt{-g} f(R)$ and we apply the variational principle $\delta \int d^4 x \sqrt{-g} f(R)=0$. This gives:
\bea
\delta_{g^{\mu\nu}}S&=&\int_\Omega d^4x\left[ \delta(\sqrt{-g})f( R)+\sqrt{-g}\,\delta (f( R)) \right]
\nonumber\\
&=&\int_\Omega d^4x\left[ -\frac{1}{2}\sqrt{-g}\,g_{\mu \nu}f( R)\,\delta g^{\mu \nu}+\sqrt{-g}\,f'( R)\delta (g^{\mu \nu}R_{\mu \nu}) \right]\\
\label{eq:first_variation}
&=&\int_\Omega d^4x\sqrt{-g}\left[  -\frac{1}{2}\,g_{\mu \nu} f( R)\,\delta g^{\mu \nu}+f'( R)\delta g^{\mu \nu}R_{\mu \nu} +f'( R)g^{\mu \nu}\delta R_{\mu \nu}  \right]\nonumber
\eea
where  $f'( R)\equiv \frac{d f( R)}{dR}$ and
\begin{equation}
\delta \sqrt{-g}=-\frac{1}{2}\sqrt{-g}\,g_{\mu \nu}\delta g^{\mu \nu}.
\end{equation}
The last term in eq.~\eqref{eq:first_variation} gives rise to the boundary effects because $\delta R_{\mu \nu}$ contains $(\delta \partial g)_{\partial \Omega}$ that it is nonzero.
First of all, let's notice that:
\begin{equation}
\delta R_{\mu \nu}=\nabla_\alpha\delta{\Gamma_{\mu \nu}}^\alpha-\nabla_\mu \delta{\Gamma_{\alpha \nu}}^\alpha.
\end{equation}
where $\Gamma_{\mu \nu}^{\alpha}$ are the usual Christoffel symbols constructed from $g_{\mu \nu}$.
Now, let's rewrite our relation in the local inertial frame where $\Gamma=0\Rightarrow \nabla\rightarrow \partial$ and the metricity condition becomes $\partial_\alpha g_{\mu \nu}=0$. In this way we obtain:
\begin{eqnarray}
\delta {\Gamma_{\mu \nu}}^\alpha&=&\frac{1}{2}\,g^{\alpha \rho}\left( \partial_\mu \delta g_{\nu \rho}+\partial_\nu \delta g_{\rho \mu}-\partial_\rho \delta g_{\mu \nu} \right)\\
\delta {\Gamma_{\alpha \nu}}^\alpha&=&\frac{1}{2}\,g^{\alpha \rho}\partial_\nu \delta g_{\rho \alpha}
\end{eqnarray}
and
\begin{equation}
\label{eq:second_variation}
\left( g^{\mu \nu}\delta R_{\mu \nu} \right)_{\Gamma=0}=\partial^\rho \partial^\nu \delta g_{\rho \nu}-g^{\alpha\rho}\partial_\mu \partial^\mu \delta g_{\alpha \rho}.
\end{equation}
If we release the inertial frame hypothesis, we have to replace $\partial$ with $\nabla$, so that eq.(\ref{eq:second_variation}) becomes:
\begin{eqnarray}
\label{eq:third_variation}
g^{\mu \nu}\delta R_{\mu \nu} &=&  \nabla^\rho \nabla^\nu \delta g_{\rho \nu}-\nabla_\mu \nabla^\mu \delta \left( g^{\alpha\rho}g_{\alpha \rho}\right)\nonumber
\\ 
&=& \nabla_\mu\left[ g_{\alpha \beta} \nabla^\mu\delta g^{\alpha \beta}-\nabla_\nu\delta g^{\mu \nu} \right]
\end{eqnarray}
where we used $g^{\alpha \beta}\,\delta g_{\alpha \beta}=-g_{\alpha \beta}\,\delta g^{\alpha \beta}$, $\delta g_{\alpha \beta}=-g_{\alpha \rho}\,g_{\beta \sigma}\,\delta g^{\rho \beta}$ and the metricity condition $\nabla g=0$.
In other word, the last term in (\ref{eq:first_variation}) becomes:
\begin{eqnarray}
\int_\Omega & d^4x&\,\sqrt{-g}\,f'( R) \nabla_\mu \left[g_{\alpha \beta} \nabla^\mu\delta g^{\alpha \beta}-\nabla_\nu\delta g^{\mu \nu}\right] \nonumber
\\
&=& \int_\Omega d^4x\,\sqrt{-g}\,\nabla_\mu\left[ f'( R)(g_{\alpha \beta} \nabla^\mu\delta g^{\alpha \beta}-\nabla_\nu\delta g^{\mu \nu})\right] \nonumber
\\
&-&\int_\Omega d^4x\sqrt{-g}\,\{\nabla_\mu\left[ \left( \nabla^\mu f'( R) \right)\,g_{\alpha \beta}\,\delta g^{\alpha \beta} \right] -\nabla_\alpha \nabla^\alpha f'( R)\,g_{\mu \nu}\,\delta g^{\mu \nu}\}
 \nonumber
\\
&+&\int_\Omega d^4x\sqrt{-g}\,\{\nabla_\nu\left[ \left( \nabla_\mu f'( R) \right)\,\delta g^{\mu \nu} \right]-\nabla_\nu \nabla_\mu f'( R)\,\delta g^{\mu \nu}\}.
\end{eqnarray}
It is important to note that second integral and the fourth one do not contribute to the variation; in fact they can be changed in two flux integrals evaluated on the boundary $\partial \Omega$, where $\delta g=0$. On the other hand, the third integral and the fifth one give a relevant contribute to the variation, that appears as follows:
\begin{equation}
\label{eq:variation1}
\int_\Omega d^4x\,\sqrt{-g}\,\delta g^{\mu \nu}  \left[ R_{\mu \nu} f'( R)-\frac{1}{2}\,g_{\mu \nu}f( R)+ g_{\mu \nu}\Box f'( R)-\nabla_\mu \nabla_\nu f'( R) \right]
\end{equation}
where $\Box\equiv \nabla_\alpha \nabla^\alpha$. Here we may add the variation of the material action which gives the standard condition
\begin{equation}
\label{eq:variation2}
\int_\Omega d^4x\,\sqrt{-g}\,\frac{1}{2}\,T_{\mu \nu}\,\delta g^{\mu \nu}, 
\end{equation}
in order to obtain all the terms proportional to $\delta g^{\mu \nu}$. There is still another term that contributes to the variation; this one can be rewritten as a flux integral as follows:
\begin{equation}
\label{eq:border_variation}
\int_{\partial\Omega}dS^\mu\sqrt{-g}\,f'( R)\left[ g_{\alpha \beta} \nabla^\mu \delta g^{\alpha \beta}-\nabla_\nu\delta g^{\mu \nu} \right]\equiv \delta_g S_{b}.
\end{equation}
In this way, we have obtained a system of fourth-order differential bulk equations that must determine 10 components of the symmetric tensor $g_{\mu\nu}$. Therefore, we are allowed to fix 40 initial conditions, in order to have a well defined solvable problem. We already fixed 20 of them by requiring that $\delta g_{\mu\nu}|_{\partial\Omega}=0$ so we are still left with 20 conditions, that are not enough to completely eliminate the boundary terms. In fact, in eq.(\ref{eq:border_variation}), we have  80 degrees of freedom.
In GR ($f'( R)=0$) the boundary contribution can be eliminated by adding the well-know York-Gibbons-Hawking action:
\begin{equation}
S_{YGH}=\int_\Omega d^4x\sqrt{-g}\,\nabla_\mu V^\mu_{YGH} =\int_{\partial \Omega}d^3\xi\sqrt{|h|}\,2K
\end{equation}
where $h_{\alpha\beta}\equiv g_{\alpha\beta}+\epsilon\,\eta_\alpha\eta_\beta$ in the induced metric on the boundary, and $K\equiv h^{\mu\nu}K_{\mu\nu}=2h^{\mu\nu}\nabla_\mu\eta_\nu$ is the extrinsic curvature of the boundary hypersurface.
Therefore, in our case a possible boundary term is:
\begin{equation}
S_{B}=\int_{\partial \Omega}d^4x\sqrt{-g}\,\nabla_\mu\left[f'( R)\,V^\mu_{YGH}\right].
\end{equation}
By varying with respect to $g$, we obtain that:
\begin{equation}
\label{eq:total_boundary_variation}
\delta_gS_{B} =  \int_{\partial\Omega}d^3\xi\sqrt{|h|}\,f'( R)\eta^\mu h^{\nu\alpha}\partial_\mu\delta g_{\nu\alpha} +\int_{\partial\Omega}dS_\mu\sqrt{-g}\,2K\,f''( R)\,g^{\mu\nu}\delta R_{\mu\nu}.
\end{equation}
The first integral of eq.(\ref{eq:total_boundary_variation}) exactly eliminate the contribution of eq.(\ref{eq:border_variation}). In addition, $R_{\mu\nu}$ is a symmetric tensor, so that the  requirement $\delta R_{\mu\nu}=0$  on the boundary can be used to impose the additional 20 initial conditions  allowing us to completely solve the bulk equations:
\begin{equation}
\label{eq:fieldEq}
R_{\mu \nu} f'( R)-\frac{1}{2}\,g_{\mu \nu}f( R)+g_{\mu \nu}\Box f'( R)-\nabla_\mu \nabla_\nu f'( R)=-\frac{1}{2}T_{\mu\nu}.
\end{equation}
Before concluding this section, let us notice that $\left(g^{\mu\nu}\delta R_{\mu\nu}\right)_{\partial \Omega}=\left( \delta R\right)_{\partial\Omega}$; in other words, the total contributions on the boundary can be rewritten as a scalar degree of freedom, using the well-know equivalence between $f( R)$ gravity and scalar-tensor theory \cite{Dyer:2008hb}.
\vskip 2truecm
%
%
%%%%%%%%  SECTION III %%%%%%%%%%%%%%%
%
%
\section{Perturbed Lagrangian}
In this Section we investigate a perturbation of the general relativity solution in a purely matter dominated Friedmann universe. In particular we'll obtain a modified expression for the expansion parameter $a$.
\\
To this end, let us consider a generic and unknown Lagrangian for a modified theory of gravity which depends only on the Ricci scalar $R=g^{\mu\nu}R_{\mu\nu}$, so we can write $\mathcal{L}=f( R)$. As we saw in the previous section the field equations are given by (\ref{eq:fieldEq}). Because we are restricting to Friedmann-Lema\^itre-Robertson-Walker (FLRW) models, with only one function to determine, we consider the trace of these equations:
\begin{equation}
\label{eq:Trace}
R \, f'( R)-2f( R)+3\,\Box f'( R)=-\frac{1}{2}\,T.
\end{equation}
In general, $f( R)$ is not specified, therefore it is possible to consider its power series expansion instead of the exact form:
\begin{equation}
\label{eq:powSerExp}
f( R)=\sum_{n=0}^\infty c_n R^n \, .
\end{equation}
From (\ref{eq:powSerExp}), it directly follows that
\begin{equation}
f'( R)=\sum_{n=1}^\infty n\,c_n\,R^{n-1}.
\end{equation}
In this way, equation (\ref{eq:Trace}) can be rewritten as follows:
\begin{equation}
\left( \sum_{n=1}^\infty   n\,c_n\,R^{n-1} \right)R   -2   \sum_{n=0}^\infty c_n\,R^n+ 3\Box \left( \sum_{n=1}^\infty n\,c_n\,R^{n-1} \right) = -\frac{1}{2}\,   T 
\end{equation}
so we obtain:

\begin{equation}
\sum_{n=1}^\infty c_n\left[ (n-2)R^n+3\,n\,\Box R^{n-1} \right]=-\frac{1}{2}\,T+2\,c_0.
\end{equation}
At this point we assume that there isn't a cosmological constant term, i.e. that $c_0=0$, because our purpose is to explain the apparent acceleration of the universe as an effect due to the Lagrangian's higher order terms without introducing any kind of dark energy. Furthermore, let's divide all of the equation by $c_1=-\frac{1}{2\chi}$ and let's define $C_n=\frac{c_n}{c_1}$, so:
\begin{equation}
\label{eq:equations}
-R=\chi\,T, \;\;\;\;\;\;\;\;\; \;\;\;\;\;\;\;\;\;\;\;\;\text{if } \;\;\;\;\;\;\;n=1
\end{equation}
\begin{equation}
-R+6\,C_2\,\Box\,R=\chi\,T, \;\;\;\;\text{if } \;\;\;\;\;\;\;n=2
\end{equation}
and so on. We stop our expansion because we assume that all of the relevant corrections to general relativity can be treated as a first order perturbative correction. Having this consideration in mind, let us consider a flat FLRW metric:
\begin{equation}
ds^2=dt^2-a^2(t)\left[ dx^2+dy^2+dz^2 \right].
\end{equation}
In such a way, we obtain $\Box R =\ddot R+3\,\frac{\dot a}{a} \dot R$; so eq.~(\ref{eq:equations}) becomes:
\begin{equation}
-R=\chi\,T, \;\;\;\;\;\;\;\;\;\;\;\;\;\;\;\;\;\;\;\;\;\;\;\;\;\;\;\;\;\;\;\;\;\;\;\;\text{if } \;\;\;\; n=1
\end{equation}
\begin{equation}
-R+6\,C_2\,\left(\ddot R+3\,\frac{\dot a}{a} \dot R\right)=\chi\,T, \;\;\;\;\text{if } \;\;\;\;\; n=2.
\end{equation}
Now, because we are interested in finding the perturbative solution of a GR solution for a purely matter dominated Friedmann model, let us expand our function as $a(t)\approx a_0(t)+C_2\,a_1(t)$, considering that $T = \rho_0\,a^{-3}\approx \rho_0\, a_0^{-3}\left( 1-3\,C_2 \frac{a_1}{a_0} \right)$
\begin{eqnarray}
\label{eq:unperturbedFRW}
&-&R\left[a_0(t)\right]=\frac{\chi\,\rho_0}{a_0^3}\\
\label{eq:firstCorrectionFRW}
&-&R\left[ a(t) \right]+6\,C_2\,\left\{ \ddot R\left[ a(t) \right]+3\,\frac{\dot a(t)}{a(t)}\dot R\left[ a(t) \right] \right\}=\chi\,T.
\end{eqnarray}
The solution of eq.~(\ref{eq:unperturbedFRW}) is given by $a_0(t)=\left( 1+\frac{\sqrt{3\chi\rho_0}}{2}\,t \right)^{2/3}$. Now we want to find the expression of $a_1$ by expanding eq.~(\ref{eq:firstCorrectionFRW}) as follows:
\begin{equation}
\label{eq:solvenda}
-R\left[ a_0+C_2\,a_1 \right]+6\,C_2\,\left\{ \ddot R\left[ a_0+C_2\,a_1 \right]
+ 3\,\frac{\dot a_0+C_2\,\dot a_1}{a_0+C_2\,a_1}\,\dot R\left[ a_0+C_2\,a_1 \right]  \right\}=\frac{\chi\,\rho_0}{a_0^3}\left( 1-3\,C_2\frac{a_1}{a_0} \right)
\end{equation}
we obtain
\begin{equation}
-R\left[ a_0 \right]-C_2\,a_1\,\frac{\partial R\left[ a_0 \right]}{\partial a}+6\,C_2\,\left\{ \right. \ddot R\left[ a_0\right]
+3\,\frac{\dot a_0}{a_0}\,\dot R\left[ a_0 \right]  \left.  \right\}=\frac{\chi\,\rho_0}{a_0^3}\left( 1-3\,C_2\frac{a_1}{a_0} \right) \, ,
\end{equation}
that is to say
\begin{equation}
a_1=\left\{6\,\ddot R\left[ a_0\right]+18\,\frac{\dot a_0}{a_0}\,\dot R\left[ a_0 \right]\right\}\left(\frac{\partial R\left[ a_0 \right]}{\partial a}-3\,\frac{\chi\rho_0}{a_0^3}\right)^{-1}.
\end{equation}
At this step, remembering that $R(t)=-6\left\{ \left( \dot a/a \right)^2+\ddot a/a \right\}$ and inserting the expression for $a_0$, we finally find:
\begin{equation}
a_1(t)=\frac{3\,\chi\rho_0}{2\left( 1+\frac{\sqrt{3\chi\rho_0}}{2}\,t \right)^{4/3}}.
\end{equation}
and then:
\begin{align}
\label{eq:solution1}
a(t)&\approx \left( 1+\frac{\sqrt{3\chi\rho_0}}{2}\,t \right)^{2/3}+C_2\frac{3\,\chi\rho_0}{2\left( 1+\frac{\sqrt{3\,\chi\rho_0}}{2}\,t \right)^{4/3}}\nonumber\\
&=\left( 1+\frac{3\,h}{2}\,t \right)^{2/3}+C_2\frac{9\,h^2}{2\left( 1+\frac{3\,h}{2}\,t \right)^{4/3}}
\end{align}
that is the most general expression of the expansion parameter in our model and directly reduces to general relativity solution for $C_2=0$, where we defined $h\equiv\frac{\sqrt{3\,\chi\rho_0}}{3}$.
\\
Differently from the usual GR normalization, the perturbed scale factor as written in eq.~\eqref{eq:solution1} is not equal to one at the present time. Nevertheless we can recover the usual normalization by redefining $a( t)$ as $a( t)-\frac{9C_2 h^2}{2}$.  This is possible by taking into account Eq. (30). In fact, let us consider the differential equation for the first-order correction $a_1$ following from Eq. (30). Using the explicit form of the zeroth-order solution, $a_0(t) =(1+ 3ht/2)^{2/3}$, and using the definition $h^2= \chi \rho_0/3$, we find that all terms proportional to $a_1$ cancel each other, and we are left with an equation for $a_1$ containing only its time derivatives $\dot a_1$ and $\ddot a_1$. Hence, we can safely subtract a constant term from the solution (33), and the result is still a viable solution for $a_1(t)$. 
\\
From now on, we will then refer to the following expression for the scale factor:
\begin{equation}
\label{eq:solution2}
a(t)=\left( 1+\frac{3\,h}{2}\,t \right)^{2/3}+C_2\frac{9\,h^2}{2\left( 1+\frac{3\,h}{2}\,t \right)^{4/3}}-\frac{9\,C_2 h^2}{2},
\end{equation}
which is automatically normalized to 1 nowadays (i.e., at $t=0$). Eq. (35) represents a very interesting dependence of the scale factor $a(t)$ on the time $t$ and we note that,
because of the normalization $a_0(0)=1$ and $a_1(0)=0$, the perturbative approach we are using is well grounded. It is true that, since $a_0$ decreases towards the past while $a_1$increases, if we consider a small but finite $C_2$, then the perturbative correction $|C_2\frac{a_1}{a_0}|$ in the past was larger than today. However, thanks to the chosen normalization $C_2{a_1(0)}/{a_0(0)}=0$,  we can still expect that the perturbative approach is valid in a given range of time, depending on the values of $h$ and $C_2$.
\\
In the following sections we will calculate some fundamental parameters like the Hubble function and the acceleration. In particular, the last section is dedicated to the comparison with supernovae Ia Union2 data, in order to obtain an experimental estimate of $C_2$ by a best-fit procedure.
\vskip 2truecm
%
%
%
%%%%%%%%%%%%% SEZIONE IV %%%%%%%%%%%%%%%
%
%
%
\section{Hubble parameter}
In the previous section we have found the solution for $a(t)$, that contains two free parameters $h$ and $C_2$. In general relativity $h$ is just the Hubble parameter evaluated today. However, in the perturbed approach, this interpretation is no longer viable: in fact, remembering that the Hubble function is related to the expansion rate by
\begin{equation}
\label{eq:Hubble_function}
H(t)=\frac{\dot a(t)}{a(t)}\approx \frac{\dot a_0}{a_0}\left[ 1+C_2\,\left( \frac{\dot a_1}{\dot a_0}-\frac{a_1}{a_0} \right) \right]
\end{equation}
we obtain
\begin{equation}
\label{eqxx}
H(t)= \frac{2 h}{2+3 h t}+\frac{9\,C_2\,h^3}{2}\left[ \left( 1+\frac{3}{2}\,h\,t \right)^{-5/3} -3\left( 1+\frac{3}{2}\,h\,t \right)^{-3}\right].
\end{equation}
So the Hubble constant is:
\begin{equation}
\label{eq:Hubble_constrain}
H_0\equiv H(0)=h-9\,C_2\,h^3.
\end{equation}
At this point, using the observed value, say $H_0=67$ km/s Mpc$^{-1}$, we are able to determine a relation among $C_2$ and $h$ by inverting (\ref{eq:Hubble_constrain}), i.e.
\begin{equation}
\label{eq:C2ofh}
C_2=\frac{h-H_0}{9\,h^3}.
\end{equation}
This relates $C_2$ to $h$ by $H_0$ and allows us to rewrite eq.~(35) as
\begin{equation}
\label{eq:solution2}
a(t)= \left( 1+\frac{3}{2}\,h\,t \right)^{2/3}+\frac{h-H_0}{2\,h}\left[\left( 1+\frac{3}{2}\,h\,t \right)^{-4/3}-1\right].
\end{equation}

This expression contains the only parameter $h$ and we notice that, as $C_2$ approaches  $0$, we have $h\rightarrow H_0$. By the way we stress once again that this interpretation falls down in the perturbed model where $h$ is just a parameter fixed by the  initial condition $\dot a(0)=H_0$.
\\
In the next section, we shall use our solution eq.~(40) in order to study the apparent acceleration of the universe. 
\vskip 2truecm
%
%
%%%%%%%%%%% SECTION V %%%%%%%%%%%
%
%
\section{Acceleration and Luminosity distance}
In order to study the acceleration properties of our model, let us compute the second  derivative of $a(t)$:
\begin{equation}
\label{eq:acceleration}
\ddot a(t)=-\frac{h^2}{2\,\left( 1+\frac{3}{2}\,h\,t \right)^{4/3}}+\frac{h-H_0}{2}\frac{7\,h}{\left( 1+\frac{3}{2}\,h\,t \right)^{10/3}}.
\end{equation}
The previous expression, when evaluated at the present time, becomes equal to $(\ddot a/a)_\text{today}$ because of our choice $a(0)=1$, so, from eq.~(\ref{eq:acceleration}) we have:
\begin{equation}
\label{eq:acceleration_today}
\left(\frac{\ddot a}{a}\right)_\text{today}=\frac{h}{2}\left( 6\,h-7\,H_0 \right).
\end{equation}
Therefore  acceleration at the present epoch appears if  eq.~(\ref{eq:acceleration_today})  is greater than 0 i.e. 
\begin{equation}
h<0 \qquad \text{or}\qquad  h>\frac {7} {6} \,H_0.
\end{equation}
This analysis shows a very important and interesting feature of this perturbative model. Without introducing any kind of dark energy, an accelerated expansion of the Universe is possible even in presence of a purely matter component and in presence of a little correction in the Lagrangian.
\\
At this point, it is very interesting to check our model with the  Supernovae Ia data considering, in particular, the recent Union2 compilation \cite{union2}. To this end, let us evaluate the luminosity distance: it is well know that, in a Friedmann-Lema\^itre-Robertson-Walker spacetime, the luminosity distance $d_L$ is:
\begin{equation}
d_L(z)=\left( 1+z \right)\int_0^z\frac{dz'}{H(z')}
\end{equation}
where $H(z)$ is the Hubble function given in eq.~(\ref{eq:Hubble_function}) with the constrain eq.~(\ref{eq:Hubble_constrain}). The dependence on the redshift is obtained by considering  the expression of $z$ for a stationary and geodesic observer, i.e.
\begin{equation}
1+z=\frac{1}{a(t)}.
\end{equation}
Moreover, defining $y\equiv\left( 1+\frac{3}{2}\,h\,t \right)^{2/3}$, it is possible  rewrite last equation as:
%
%
%\textcolor{blue}
{\begin{align}
\frac{1}{1+z}=a(t)=y+\frac{h-H_0}{2h}\left( y^{-2}-1 \right)\Rightarrow
y^3-A_1(z)\,y^2+A_2=0
\end{align}}
where
\begin{eqnarray}
&A_1(z)=\frac{h-H_0}{2\,h}+\frac{1}{1+z}\nonumber\\
&A_2=\frac{h-H_0}{2\,h}.
\end{eqnarray}
Finally, by solving the third order polynomial, we find that
\begin{equation}
y(z)=\frac{1}{3}\left[ A_1+\frac{A_1^2}{A_3}+A_3 \right]
\end{equation}
or equivalently
\begin{equation}
t(z)=\frac{2}{3\,h}\left[y(z)^{3/2}-1\right]
\end{equation}
with 
\begin{equation}
A_3^3(z)=A_1^3+\frac{3}{2}\left( \sqrt{81\,A_2^2-12\,A_2A_1^3}-9A_2 \right).
\end{equation}
Once  this relation is found, we can insert it into definition of $d_L$ and, by numerical integration, we are able to evaluate the so-called distance modulus 
\begin{figure}[h!]
\centering
\includegraphics[scale=0.5]{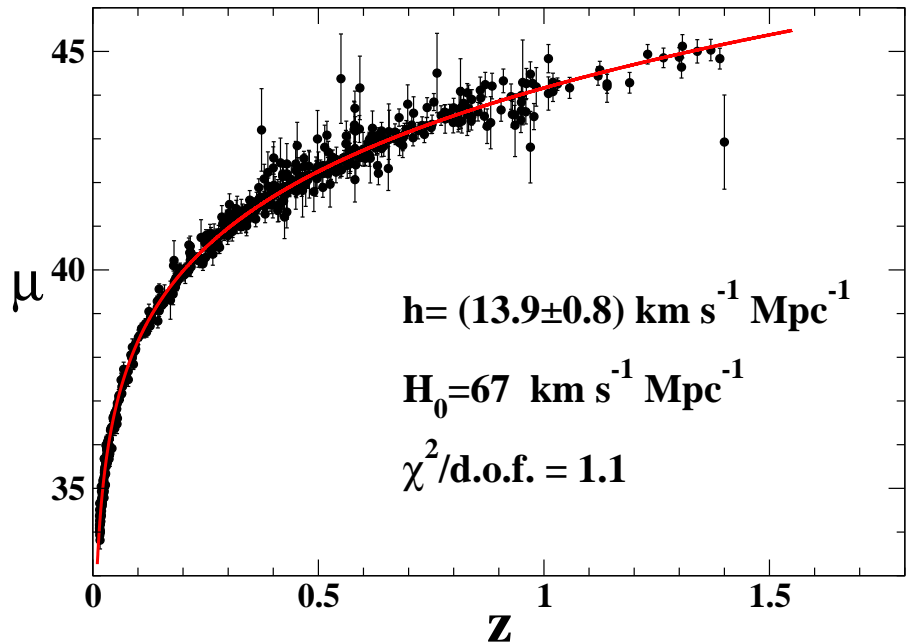}
\caption{The Hubble diagram of the Union 2 dataset. The plot illustrates the best-fit result with the only parameter h. We have h=$(13.9 \pm 0.8)$ km/s Mpc$^{-1}$ with $ {\chi^2 } / \text{d.o.f.} =1.1$. We have considered for the Hubble constant the value ${H_0=67}$ km/s Mpc$^{-1}$. }
\label{figure2}
\end{figure}
\begin{equation}
\label{mu}
\mu(z,h)=5\log_{10} \left[\frac{d_L(z, h)} {1Mpc}\right] +25\, .
\end{equation}
We have in principle one free parameter $h$, since we will use for the Hubble constant the recent value $H_0= 67$ Km/s Mpc$^{-1}$ given by Planck data. It is  then possible to fit the experimental data $\mu^\text{obs} (z_i) \pm \Delta \mu(z_i)$, where  $\Delta \mu(z_i)$ is the relative error of the modulus distance with respect to the $i-th$ Supernova Ia, by means of a $\chi^2$ analysis with
\begin{equation}
\label{chi2}
\chi^2 = \sum_{i=1}^{557} \left[ \frac{\mu^\text{obs} (z_i) - \mu(z_i, h)} {\Delta \mu(z_i)} \right]\, .
\end{equation}
In Fig. 2 we plot the distance modulus vs redshift in the Hubble diagram. Minimizing the $\chi^2$ expression we find the best -fit value $h=(13.9 \pm 0.8)$ Km/s Mpc$^{-1}$ with $\chi^2/ \text{d.o.f.} = 1.1$. The best-fit red curve is superimposed to the Union 2 data set (with error bars). It is then interesting to see that the model considered here is able to reproduce  the corresponding best-fit results for a homogeneous $\Lambda$CDM model in Friedman-Lemaitre-Robertson-Walker metric, without introducing the cosmological constant.

Moreover, the value we found for $C_2$ is $-(2.2\pm0.4)\times10^{-3}$ (km/s Mps$^{-1})^{-2}$. It is important to stress  that this best-fit value corresponds to a decelerated expansion of the Universe and that, with these values of $h$ and $C_2$, the condition required for the validity of our perturbative expansion is satisfied in the appropriate redshift range.

%
%
%***************************************************************************************
%
%
%
\section{Conclusion}
The cosmic acceleration produced an intriguing shock to cosmologists in 1998. Cosmological constant, dark energy, backreaction of inhomogeneities has been invoked in General Relativity. It is also possible to fit supernovae data with different models, for example LTB models in \cite{Cosmai:2013iga} or LTB-anisotropic model of the Universe \cite{Fanizza:2014tua}.
 A theory of modified Einstein gravity could be an explanation. 
In this research we have investigated  a perturbative approach based on a FLRW metric in $f(R)$ gravity and we have studied a model  which in general is able to describe data about the supernovae Ia. This model mimics a cosmological evolution consistent with observations. 
\\
Among the many forms of the function $f(R)$ present in the literature, here we discuss an expansion of f(R) in power series of $R^n$. Capozziello {\it et al.} \cite {{capofinale},{capofinale2}} have introduced an action with a term $f(R) \sim R^n$ and they have shown that this lead to an accelerated expansion for $n \simeq 3/2$.

We consider a second order expansion of the cosmological parameter $a(t)$. We have found an approximate expression given by eq.~(\ref{eq:solution1}) that contains two parameters but, taking into account the observed value of the Hubble constant, it is possible to reduce ourself to the case of a single-parameter model, see eq.~(\ref{eq:solution2}). Notice that we used the Hubble parameter's value determined by Planck, in order to (possibly) reduce the tension of the $H_0$ determination between the CMB observations and the SNIa ones. In this context there is no need  of  introducing a dark energy component in order to explain supernovae data: in fact, fitting the data released by Union 2 we have found  that the experimental points in the Hubble diagram can  be accurately described also by   model presented in this paper.  \\
In conclusion we want to stress that this work offers a possible explanation of the apparent acceleration of the Universe as a consequence of a dynamical approach in which we consider a perturbed general relativity solution, based on a modified gravitational theory, without introducing cosmological constant and/or dark energy. Therefore it is possible that the usually claimed acceleration effect is not the manifestation of an increase of the expansion velocity, but rather the first signal of a  gravitational Lagrangian different from the Einstein-Hilbert Lagrangian. From a philosophical point of view the core of the problem is that we have a limited number of cosmological tests available, in order to discriminate among different theories candidate to explain the observed Universe.
\\
Keeping in mind that there are numerous possibility for $f(R)$, we do not forget that our choice is one of the most simple to consider. No doubts, the details must be more complicated that the model discussed here.  The study if this model may also provide some specific effects that could discriminate between other possibilities. This will be done in a future work. 
\\
\\
\section*{Aknowledgements}

The authors would like to thank M. Gasperini for useful discussions. This work is supported by the research grant ``Theoretical Astroparticle Physics'' No. 2012CPPYP7 under the program PRIN 2012 funded by the Ministero dell'Istruzione, Universit\`a e della Ricerca (MIUR). This work is also supported by the italian Istituto Nazionale di Fisica Nucleare (INFN) through the ``Theoretical Astroparticle Physics'' (TASP) project.
%
%\section*{References}
%
 
\end{document}